\documentclass[twoside,letterpaper]{soups} 
\usepackage{caption}
\usepackage{graphicx}
\usepackage{booktabs}
\usepackage{url}

\begin{document}
%

\title{Online Social Media and Police in India: \\Behavior, Perceptions, Challenges}%
%
%
%
%
%

\numberofauthors{1} 
\author{
\alignauthor
Niharika Sachdeva, Ponnurangam Kumaraguru \\ 
\affaddr{ Cybersecurity Education and Research Centre (CERC)}\\ 
\affaddr{Indraprastha Institute of Information Technology - Delhi} \\ 
\email {niharikas@iiitd.ac.in, pk@iiitd.ac.in}
}

%
%
%


\maketitle
\begin{abstract}
Police agencies across the globe are increasingly using Online Social Media (OSM) to acquire intelligence and connect with citizens. Developed nations have well thought of strategies to use OSM for policing. However, developing nations like India are exploring and evolving OSM as a policing solution. India, in recent years, experienced many events where rumors and fake content on OSM instigated communal violence. In contrast to traditional media (e.g. television and print media) used by Indian police departments, OSM offers velocity, variety, veracity and large volume of information. These introduce new challenges for police like platforms selection, secure usage strategy, developing trust, handling offensive comments, and security / privacy implication of information shared through OSM. Success of police initiatives on OSM to maintain law and order depends both on their understanding of OSM and citizen's acceptance / participation on these platforms. 

This study provides multidimensional understanding of behavior, perceptions, interactions, and expectation regarding policing through OSM. First, we examined recent updates from four different police pages -- Delhi, Bangalore, Uttar Pradesh and Chennai to comprehend various dimensions of police interaction with citizens on OSM. Second, we conducted 20 interviews with IPS officers (Indian Police Service) and 17 interviews with citizens to understand decision rationales and expectation gaps between two stakeholders (police and citizens); this was followed up with 445 policemen surveys and 204 citizen surveys. We also present differences between police expectations of Indian and police departments in developed countries.

\end{abstract}

\section{Introduction}
Online Social Media (OSM) is extensively used by first responder such as police to acquire intelligence and offer better services to citizens~\cite{Thomas-Heverin:2010uq, Inforgraphicpoliceuse, Silva:2013ys}. Police departments across the world have their own profiles on Twitter to communicate with people e.g. Spanish police, Seattle police and Chicago police~\cite{Post:2013ve, Seattlepolice, Silva:2013ys}. They use OSM  to provide better services such as beat meetings (neighborhood police interaction session), business, and missing person cases. In UK, Greater Manchester Police started a program on OSM where citizens could join police on the beat and blog / tweet about their experiences~\cite{GMPonsocialmedia}. 
Similarly, Seattle police launched @GetYourBikeBack program on OSM to help owners get their lost bikes~\cite{getyourbikeback:2013zr}. A large number of citizens follow police departments on Twitter and Facebook to obtain real-time crisis updates. For instance, the news of Boston explosions broke on Twitter more than 10 minutes before the national media started reporting about it~\cite{TwitteristhenewPoliceScanner:2013vn}.

\begin{figure}[!htb] \scriptsize
\captionsetup{font=scriptsize, labelfont=bf, textfont=bf}
\centering
{\includegraphics [width=\linewidth]{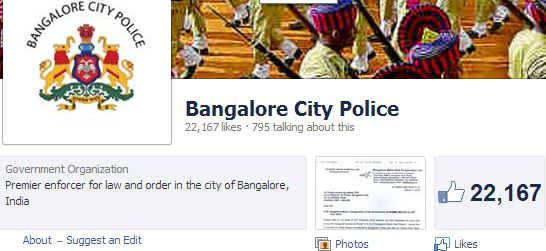}}
\caption{Bangalore City Police page with 22,167 likes. {https://www.facebook.com/blrcitypolice}}
\label{fig:policebangalore}
\end{figure}

OSM exhibits great potential to transform policing experience, increase transparency, provide timely information and offer virtual interaction with public~\cite{Silva:2013ys, SocialMediaPoliceUK, FBIsocialmedia}. OSM use is not just restricted to developed countries but also spreads to developing countries like Pakistan and South Africa~\cite{Inforgraphicpoliceuse}. In this study, we focus on India as a case study to understand use of OSM for policing in developing nations. India is expected to have world's largest Facebook population in 2016 with a penetration rate of 7.7\%~\cite{Indiafacebookfact}. OSM has played an instrumental role in spreading rumors and mobilization of masses during crises in India such as Muzzafarnagar riots (UP, India) and Assam disturbance~\cite{Hindustan-Times:2013cr, Hindustan-Times:2013nx, PTIAssamriots:2013}. 
Ineffective communication with public proved a great hindrance for police departments in such events~\cite{Accenture-research:2012fk}.This has opened up discussions regarding use of OSM by police forces. In this study, we explore some popular city police and traffic police pages in Delhi, Bangalore, Hyderabad, and Chennai. Figure~\ref{fig:policebangalore} shows Bangalore City Police Facebook page. 

OSM introduced various challenges for police which might impede its adoption~\cite{SocialMediaPoliceUK, John-Carlo-Bertot:2012fk, Wigand:2010vn}. Police use of OSM obstructed the process of natural justice and lead to embarrassment of departments. A UK cop posted a picture on his account with seized drugs but that affected the trial of the case. Some OSM friends to influence jury misused the wall post. Similarly, a picture of cop sleeping on duty instigated questions pondering whether the cops were really fit. Personal information like occupation details filled casually on OSM caused problems for the officers. An officer described himself as a ``super hero / serial killer'' in the occupation field of OSM which was used in court of law to prove that officer used excessive force against the accused~\cite{ALFANO:2006uq, FBIsocialmedia}. The UK police profiles faced criticism for following celebrity accounts Miranda Hart (comedian), Rihanna (pop singer) and shopping brands including Victoria's Secret (lingerie company). One of the senior officer responsible for UK police social media said ``I think there needs to be a better understanding of senior managers as to where does this [OSM] fit into the overall strategy of what we are trying to do.'' Surveys across the world analyze the various activities of police on OSM~\cite{Gallagher:2001uq,Thomas-Heverin:2010uq, Research-commissioned-by-LexisNexis:2012fk}. Longitudinal studies represent influence of differences and changes between various police departments in US and UK Police. However, we found no work in context of developing nations that analyzed behaviors, perceptions, challenges, security and privacy concerns of police. 


Solution to these require broader assessment OSM system designs and acceptance of OSM for policing. We hope insights from our study  help develop OSM strategies and technologies for secure police interactions for developing nations like India. Citizens are important stakeholders of policing enterprise; our findings bring forth the needs and knowledge gap between police and citizens on OSM role for policing. Our contributions are:
{{\begin{itemize}
\item{We study behaviors, and perceptions of citizens and police on OSM for policing activities. Our study explores expectation gap between police and citizens for OSM use for policing. We do this through interviews of police officers and citizens.}
\vspace{-2mm}
\item{We explore various opportunities and vulnerabilities for police departments as they adopt OSM as a communication channel with citizens. Through surveys of 445 policemen and 204 citizens, we develop understanding of mass opinion for policing on OSM.}
\vspace{-2mm}
\item{We compare our results with existing studies on police in the US, to show cultural influence on OSM adoption by police.}
\end{itemize}}}

\vspace{-3mm}
\section{Related Work}
Recent studies show increase in need of OSM as a plausible resource for police forces~\cite{Denef-S.-Kaptein-N.-Bayerl-P.-S.:2011vn}. Police in developed nations have realized effectiveness of OSM in various activities such as investigation, identifying crime, intelligence development, and community policing~\cite{Denef-S.-Kaptein-N.-Bayerl-P.-S.:2011vn, IACP:2013kx, Lexis-Nexis-Risk-Solutions.:2012uq}. However, interactivity and pace at which information diffuses on OSM results in additional pressure on police departments~\cite{Denef-S.-Kaptein-N.-Bayerl-P.-S.:2011vn}. We found that police departments in developed countries have made reasonable efforts and progress to adopt OSM, whereas developing countries are still evolving skills to use OSM for policing~\cite{Inforgraphicpoliceuse}. Developing nations like India are not untouched from the influence of OSM. India has 92 million Facebook users (7.73\% of total Facebook user base) who are spread across both major (34\% of user base) and small cities (24\% of user base) of the country~\cite{Nayak:2014oq}. Various studies in Indian context showed that OSM was used to spread misinformation and public agitation during crisis events such as Mumbai terror attacks (2011), Muzzafarnagar riots and Assam disturbance(2012)~\cite{Gupta:2011kl, Kumaraguru:2013tg, The-times-of-India:2012hc}. In both the events (Assam, and Muzzafarnagar riots), panic was spread through fake images, messages, and videos on OSM~\cite{The-times-of-India:2012hc}. 

Existing studies have shown effectiveness of OSM for crises like Boston bombings, Sichuan earthquake (2008), Haiti earthquake (2009), Oklahoma grassfires (2009), and Chile earthquake (2010)~\cite{Gupta:2012ly, Gupta:2013zr, Mendoza-M.-Poblete-B.-and-Castillo-C.:2010qf, Qu-Y.-Wu-P.-and-Wang-X.:2009ve, Starbird-K.-and-Palen-L.:2011bh, Vieweg-S-Hughes-A-Starbird-K-and-Palen-L.:2010dq}. These studies demonstrate that OSM could provide critical real time information and reduce misinformation during crisis events. Researchers found that citizens used OSM for public coordination during crises. They categorized public response received during crisis on OSM and showed different communities which developed during crises~\cite{Gupta-A.-Joshi-A.-and-Kumaraguru-P:2012cr, Hughes-A.-L.-Palen-J.-Sutton-S.-Liu-and-S.-Vieweg.:2008nx}. We found few studies which analyzed police -- public use of OSM~\cite{SocialMediaPoliceUK, Thomas-Heverin:2010uq}. Studies showed that police organizations need effective communication strategy to provide timely information to various groups~\cite{Chermak-S.-and-Weiss-A.:2005ij}. These studies provide insights on different strategies and activities police perform on OSM. However, these provide little insight about police rationale and expectations behind these actions and citizen acceptance of these actions. Surveys showed that OSM introduced challenges for police officers such as fake / impostor accounts which target law enforcement agencies, security and privacy concerns, civil liabilities and resource constraints like time, and staff~\cite{IACP:2013kx, Lexis-Nexis-Risk-Solutions.:2012uq}. Another challenge was easy accessibility of OSM to malicious people, which could make sharing information with citizens a complex task~\cite{SocialMediaPoliceUK}. However, there was no in-depth analysis of the citizen and police perceptions / concerns which lead to these challenges. To best of our knowledge, this is the first study, which analyzed police and citizen behavior / expectation regarding OSM use for policing. We present expectations gaps between police and citizens on OSM use for policing. We believe the insights from our study would provide opportunities to develop better communication strategy for police and motivate technologist to build secure system designs for effective policing using OSM.

\vspace{-3mm}
\section{Overview: Indian police on OSM}

Various policing situations such as riots, bomb-blasts and informing citizens about crimes need mass attention in short span. OSM promises these conditions (vast reach with great velocity), however one needs to build and sustain networks in order to get benefit. Various police organization benefit from OSM and have considerable following of OSM pages (see Table \ref{tab:comparison}). These pages also define rules and regulations for citizens, if they wish to communicate with police using OSM. Departments adopt different strategies - Push (disseminate information), Pull (silently observe / obtain information) or Engage (interact and encourage two way communication on OSM) to interact with citizens. Not many police departments in US and UK allowed others to post (engaging strategy) on Facebook pages (See Table \ref{tab:comparison}). Contrary to this, police pages in India allowed citizens to post on Facebook pages. Our research motivation is to answer questions like: \emph{why citizens should be allowed / not allowed to post of police pages? and what were the associated security risks and apprehensions?} To best of our knowledge, this is the first study which analyzis police and citizens' reaction through interviews and a follow up survey to understand the rationale behind these decisions.

We also found that popularity of an OSM platform was not consistent across countries. Some states preferred to use Twitter whereas others used Facebook. For instance, Boston police had 2,65,800 followers on Twitter whereas the page had only 88,047 likes on Facebook. Contrary to this, New Your City police had 1,95,484 likes on Facebook and 84,830 Twitter followers. Despite police departments posting similar content on Twitter and Facebook pages. To understand OSM use by Indian police, we collected Facebook posts and Twitter status of 4 pages -- Delhi Traffic Police, Bangalore City Police, Uttar Pradesh Police and Chennai City Police. We found police pages in India were fairly new and had limited following (see Table~\ref{tab:comparison}). While searching for police pages we found that multiple police pages existed of each police department, for instance, Mumbai police had 6 accounts on Twitter. We asked them to identify legitimate page of their organization but we found that many policemen found it difficult to identify legitimate pages of police departments.

\begin{figure*}\scriptsize
\captionsetup{font=scriptsize, labelfont=bf, textfont=bf}
\centering
\begin{tabular}{ccc}
{\scalebox{0.40}{\includegraphics[scale = 0.50]{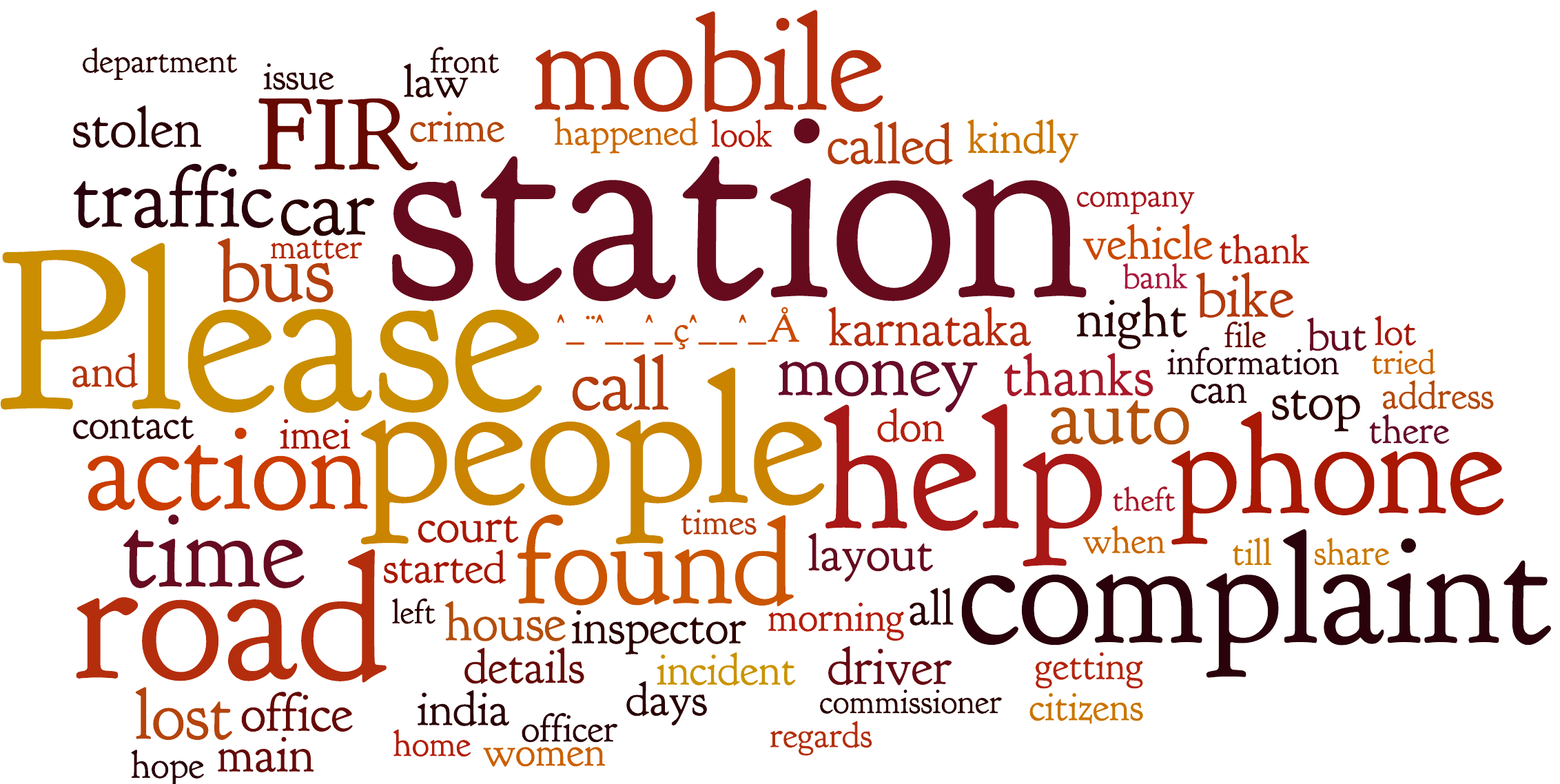}}} &{ \scalebox{0.40}{\includegraphics[scale=0.55]{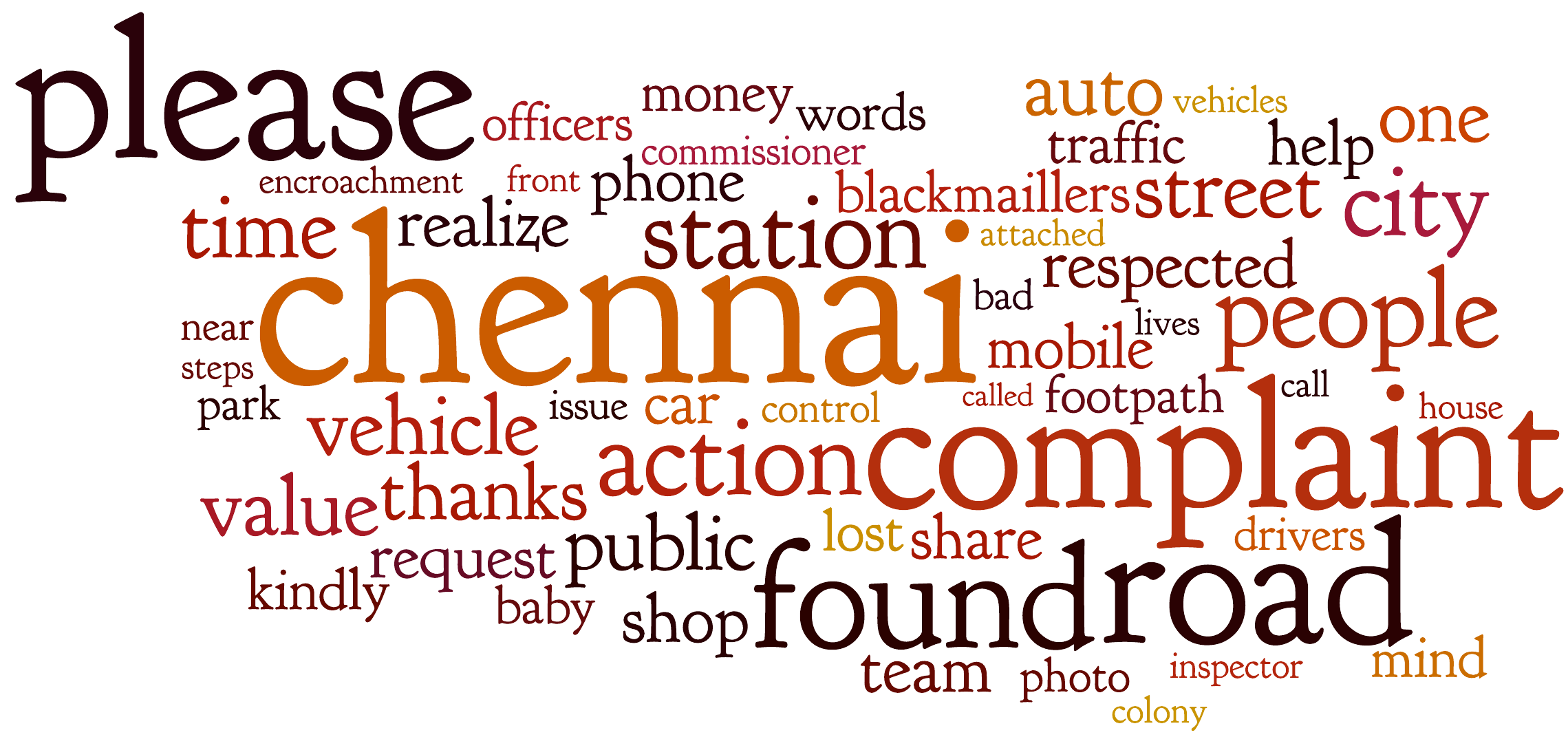}}} &{\scalebox{0.40}{\includegraphics[scale=0.55]{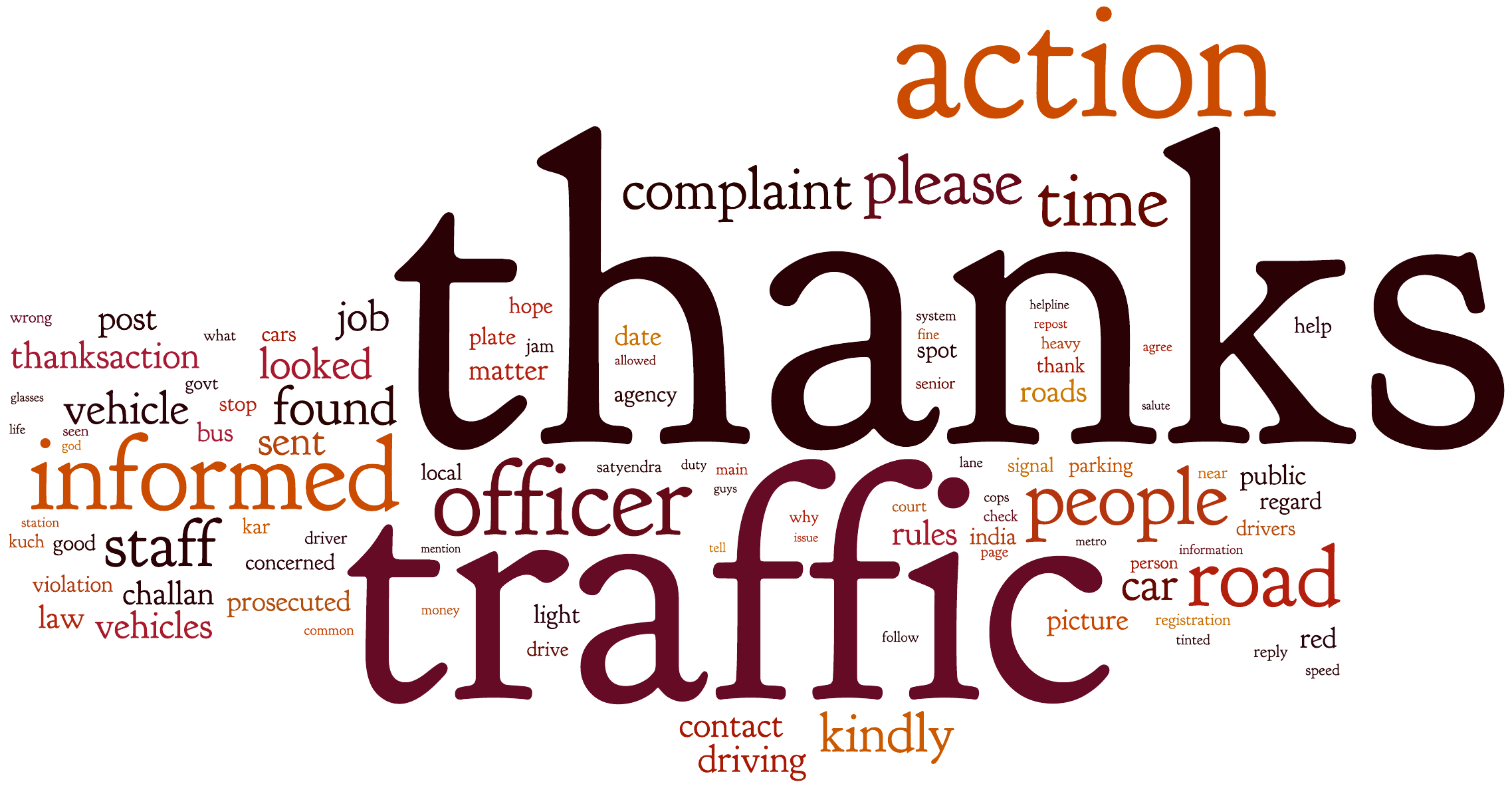}}} \\

(a)   &	(b)&	(c)\\
%
\label{wave}\end{tabular}
\vspace{-6mm}
\caption{(a) shows popular discussions on Bangalore City Police (BCP) page, (b)  shows Chennai police page, a relatively new page, was dedicated to city issues (c) shows comments posted on BCP page}\label{figure:tagclouds}
\vspace{-3mm}
\end{figure*}

Police activities mainly include maintenance of law and public order, crime prevention and detection, traffic management, and enforceing laws of the land. To perform these activities police departments require active and dynamic interaction / participation from citizens. 
We found that police in India used OSM primarily for traffic management, posting personal achievements and appreciation received by citizens. These pages also highlighted security conditions in disturbed / riots affected areas, educate citizens about current beat (patrols), and safety programs undertaken by city police. Figure~\ref{figure:tagclouds} shows tag-cloud of most frequent words of posts and comments on these pages. Bangalore City Police (BCP) page provided one such platform to report issues related to policing. Figure ~\ref{figure:tagclouds}(a) shows popular discussions on Bangalore City police page were regarding phone / mobile, finance problems, issues on roads, traffic, drivers, buses, lost objects, and First Information Reports (FIR). The comments below show complaints filed on OSM police pages. 

%


{\small{\emph{Respected Commissioner,
I wish to introduce an incident happened with us on this Tuesday, my brothers mobile got robed in bus near Madiwala.}}}


\vspace{0.2cm}

{\small{\emph{Respected Commissioner,
I wish to inform FIR No.07XX/2013 is registered in H.A.L. PS.GSC No. is PO14161301XXXXX. Registration No.: KA05HXXX, Chassis No.: ME11CK0XXX, Engine No.:1CK2011XXX, Bike Model: Yamaha R15 white color, Phone: 808886XXXX \& 808839XXXX.
}}}

\begin{table}[ht]\scriptsize
\captionsetup{font=scriptsize, labelfont=bf, textfont=bf}
  \centering
  \scriptsize
  \caption{\small{Twitter followers and Facebook (FB) likes on police pages. ** shows both FB and Twitter profiles were not verified. * shows Facebook page was not verified. ``Post'' shows if others were allowed to post on FB. ``Joined FB shows year page came in existence''}}
    \begin{tabular}{|p{3.2cm}|p{1.1cm}|p{0.8cm}|p{0.6cm}|p{0.8cm}|} \hline
    \centering
    \small
    \raggedright{Police Departments} & Followers & Likes & Post & Joined FB \\
   \hline
    \multicolumn{5}{c}{{\textbf{USA}}} \\
 \hline
 {Boston} & 265800 & 88047 & No&2010 \\
  {New York} & 84830 &195484 & No&2012\\
   {Seattle } & 51118 & {7701} & No&2010 \\
   {{Baltimore}} & 41511 & {15877} & Yes&2012 \\
    {{Metropolitan, Columbia}} & 35900 & {7297} & Yes&2008 \\ 
      \hline
    \multicolumn{5}{c}{{\textbf{UK}}} \\
    \hline
\raggedright{London*}& 139989 & {22943} & No& 2011 \\
    {{Greater Manchester*}} & 199843 & {66203} & Yes& 2011\\
  {{West Midlands*}} & 85615 & {50211} & No&2008 \\
  {Northern Ireland*} & 46639 & {14,358} & No& 2009\\
   {{Essex*}} & 46146 & {22,466} & No&2011 \\
    \hline
\multicolumn{5}{c}{{\textbf{India}}} \\
    \hline
  {{Delhi Traffic**}} & 92600 &178803 & Yes&2011 \\
    {{Bangalore City**}} & 473   & 22,167 & Yes&2011 \\
   {{Bangalore Traffic**}} & -     & {101099} & Yes&2012 \\
   {{UP Police PR**}} & 2519  &{3810} & Yes&2013\\
  {{Chennai**}} & 8     & {1605} & Yes&2013 \\
    \hline
    \end{tabular}%
  \label{tab:comparison}%
  \vspace{-3mm}
\end{table}%

People addressed various authorities like inspectors and commissioner to file complains and inquired about time constraints in which problem could be solved / addressed. To further understand citizens' view, we analyzed comments on BCP page. Mostly people reported issues in polite words -- request, please and addressed officers as ``Sir'' (see figure \ref{figure:tagclouds}(c)) but some people made rude statements about police. We found people thanked police, appreciated their work and inquired about action taken, and nearest station. Some examples below:

\vspace{0.2cm}
{\small{\emph{ Great !! absolutely great !! Reading all the complaints and statuses. I am so very happy that police is doing great job in helping people via FB as well  
}}}

\vspace{0.2cm}

{\small{\emph{ I read many posts in this page, these are very helpful for people but in real Bangalore police totally different. Its very rude 
}}}
\vspace{0.2cm}

People posted sensitive information and crime tips on Bangalore City Police page such as drug dealers active in the city, people in serious distress, and money laundering issues. 

\vspace{0.2cm}
{\small{\emph{ Drug dealer Syed. Phone number is 99002XXXXX. Please track him down. Selling ganja/ pot/ weed/ marijuana. }}}

\vspace{0.2cm}

{\small{\emph{ I Am physically and mentally tortured in this company. My health is getting effected [I] am without money since 2 months }}}

\vspace{0.2cm}


Chennai police page, a relatively new page, was dedicated to city issues. Posts included questions regarding police action taken, complaints about phone, money, shops, traffic issues on roads, and few posts were regarding blackmailers. Figure~\ref{figure:tagclouds}(b) shows frequent words in these discussions. On analyzing comments on this page, we found popular topics were about blackmailers, money, police and station. We found some violent reactions on this page, people used words like Kill, torchuring, techniques etc. Outstation people in chennai also posted their issues like:

\vspace{0.2cm}
{\small{\emph{ We the people of INDIA from West Bengal want to say that some very very dangerous groups want to kill many poor Indians for MONEY.}}}
\vspace{0.2cm}

Delhi police did not have city police page but maintained a traffic police page. We analyzed this page and found that popular issues were traffic, vehicles, people, and car. Delhi police appreciated citizen participation for informing them about routine problems and also provided assistance if needed.

\vspace{0.2cm}
{\small{\emph{ 
 Thanks, matter will be looked into and you may also contact to TI/KJC at 875087XXXX. \\
}}}

Similar to Bangalore City Police page, people gave their feedback to Delhi Traffic Police on these pages. Some people regarded these pages as eye wash whereas some people asked for status of their complaints and actions taken. For instance a citizen wrote

\vspace{0.2cm}
{\small{\emph{ You are right Irfan, but DTP can never improve. FB page is also an eye wash. }
}}\\

{\small{\emph{Delhi Police is the most efficient police force in the country under the prevailing circumstances in the country.}}}

People looked disappointed for not getting a reply / action taken against their complains in a given time frame. However, Delhi Traffic Police tried to keep citizens informed about actions taken and also posted other advisories as required. Some examples below:

\vspace{0.2cm}

{\small{\emph{
Delhi Traffic Police - NO ACTION SO FAR DAY 6.}}}
\vspace{0.2cm}

{\small{\emph{
The following vehicles/owners have been prosecuted by issuing notice on the basis of photographs on dated 08/02/2014
Vehicle no. - Notice no [XXXX]
}}}

\vspace{0.2cm}

%

Citizens also posted a variety of personal information on these pages such as phone numbers, IMEI numbers, and identity cards to report complains on these pages. Sometimes, police departments posted contact details as a reply to these post. 
People also posted irrelevant content on these pages such as publicity of a political party or general complaints against politicians. So far we analyzed messages on Facebook and Twitter to understand the policing landscape on OSM. This introduced many questions like how much information was useful for police departments, and were there some posts which could create problems for citizens. It was not clear, how OSM could help police departments to achieve their goals and what were the security / privacy / legal implications of sensitive information shared on OSM? Success of police pages could not be measured without understanding citizen's expectations and satisfaction.  We conducted interviews and surveys with both the stakeholders to gain insights about experiences with, and expectations from OSM.
%



\vspace{-3mm}
\section{Methodology}
To explore users' practices and expectations, we conducted 37 semi-structured, in-depth interviews -- 17 with citizens in Delhi, India and 20 with IPS (Indian Police Service) officers across the country. Next, we designed an online survey to gain insights about mass opinions about police on OSM. We developed a survey protocol and used it to collect 445 IPS officer responses and 204 citizen responses. In this section, we describe the structure and methodology that we used for the interviews and online studies, and present the demographics of participants. We were not required to go through an Institutional Review Board (IRB) approval process before conducting the study. However, the authors of this paper have previously been involved in studies with U.S. IRB approvals, and have applied similar practices in this study. Participants were shown consent information, which they agreed to participate in our study.


\subsection{Interviews}
Most of interviews were conducted through telephone. We recruited all participants through word-of-mouth and mailing lists dedicated for IPS officers. IPS is one among the three ``All India Service'', managed by Ministry of Home Affairs. Across the country, 3549 IPS officers serve a population of 1.27 billion i.e. 1 IPS officer for every 3,59,953 people~\cite{Ministry-of-Home-Affairs:2010uq}. Despite this, police departments represent the most omnipresent and ubiquitous body of a society. It is often considered to be the most representative and perpetual government service. Citizens are expected to consider police departments as most reachable governmental help for any kind of problem or day-to-day issues. We completed 20 individual interviews with IPS officers, lasting for about 60 minutes each. All the IPS officers whom we interviewed in the survey were Males and their age varied from 25  to 55 years and were from providing services in different states; 3 officers were serving in special branches. The other set of interviews included interaction with citizens to understand their perspective about police presence on OSM. Participants were spread across various walks of life (e.g. age group, education, and occupation). Interview sessions involved one participant at a time and were administered by one of the authors. 
All interviews were recorded, and transcribed for analysis. We used randomly generated numbers to identify the subjects in our notes so as to maintain subjects' privacy. We took participants consent before taking part in the interviews and we compensated all participants for their time and efforts.
%

Interview questions were compartmentalized into different topics such as need for OSM use for policing, how OSM has been helpful so far, understanding about policies, and what are the major hindrances in adoption of OSM for policing. Similar to IPS officers, interview questionnaire for citizens comprised of different topics such as utility of OSM in reducing the communication gap between police and citizens, ways in which citizens would like police to help through OSM, and preferred medium of communication with police. \emph{Appendix A} contains link to the entire protocol that we used for police and citizen's interview. To understand IPS officers and citizens decision and rationale for using OSM during real world events, we showed / narrated them some statuses (Twitter) / posts (Facebook). Participants could choose to retweet / share, like, comment, or ignore the posts. No personal information (name, email address, etc.) that would re-identify any subject was recorded in the interview data. Table~\ref{table:demographicsofusersintheinterviews} gives demographics of participants in interviews and surveys.
%


\begin{table}[!ht]\scriptsize
\captionsetup{font=scriptsize, labelfont=bf, textfont=bf}
\centering
\scriptsize
\begin{center}
\caption[Demographics of the participants in the surveys.]{\small{Demographics of the participants in the interviews and surveys. Values in the table are in percentage.}} \label{table:demographicsofusersintheinterviews}
\end{center}
\vspace{-1em}
\scriptsize
\begin{tabular}{| p{2.5cm}| p{1cm} | p{0.8cm} | p{0.8cm}| p{1.4 cm}|}
\hline
& Citizen Survey N=204 & Police Survey N=445 & Citizen Interview N=17 & IPS Officers Interview N=20 \\
\hline
\bf{Gender}&  && &\\
\hline
Female &40.66 & 8.98 &58.82&--\\
\hline
Male & 56.59 & 85.85 &41.18&100.00\\
\hline 
Not shared & 2.75 & 5.17 &--&--\\
\hline 
\bf{Age}& & &&\\
\hline
18 -- 24 &80.22&3.05 &29.41&--\\
\hline
25 -- 34 & 16.48&37.79 &52.94&20.00\\
\hline 
35 -- 44 & 0.55&21.36 &17.65&10.00\\
\hline
45 -- 55 & 0.55&27.93 &--&55.00\\
\hline 
55- 65 & -- & 9.87&--&5.00\\
\hline
65+ & -- & &--&10.00\\
\hline
Not shared & 2.20& -- &--&\\
\hline
\bf{Education}& & &&\\
\hline
Computer IT & 54.40&&29.41&\\
\cline{1-2 } \cline{4-4}
Teaching / Research &10.99&&23.53&\\
\cline{1-2} \cline{4-4}
Fashion Designing&10.99&&11.76&\\
\cline{1-2} \cline{4-4}
MBA&3.30&Police&5.88&IPS\\
\cline{1-2} \cline{4-4}
CA & 0.55&men&&\\
\cline{1-2} \cline{4-4}
Others &19.77&&29.42& \\
\cline{1-3} \cline{4-4} \cline{5-5}
\end{tabular}
\vspace{-3mm}
\end{table}
\vspace{-2mm}
\subsection{Survey}
We designed the survey questionnaires based on the trends observed during the interviews. These included questions about general activities on OSM for policing, concerns, scope of use, hindrances and policies. In total, we collected responses from 445 policemen. Our sample consisted of respondents from different districts in Mumbai, India. We realize that number of males policemen in our study are dominant. However, male and female ratio in our study, is representative ratio of both genders in Indian police services. This is the first such study which analyzes IPS officers and policemen usage of OSM on this scale \emph{Appendix A} contains link to the entire protocol that we used for policemen and citizen survey. Due to page limit, we could include the survey questionnaires.  The second phase of survey comprised of collecting citizens views on use of OSM for policing, concerns about police presence on OSM, and activities for which police could perform using OSM. To understand citizen's perspective of these topics, we collected 204 survey responses from residents of Delhi and Noida. Table~\ref{table:demographicsofusersintheinterviews} shows the demographics of the survey participants.


\subsection{Data Analysis}
We used thematic analysis -- a qualitative research technique, to analyze interview data. Authors sorted answers for each question in different categories. These sorted answers were coded to obtain specific numbers and these codes were used to identify themes among participant's responses. We excluded the answers where participants said they ``did not know the answer''.  We analyzed the interview and survey responses to identify various trends and categorized these to show various aspects of policing on OSM. Further, we applied various statistical test such as Fishers test to present difference between opinions and perceptions. We also used $\chi^2$ test wherever appropriate to check statistical significance. Finally, we compared IPS officers responses in our study with some existing surveys of police officers across other European and the US states, to develop comparative understanding about OSM use by police departments.

\vspace{-1mm}
\section{Results}
Effective policing through OSM requires knowledge of correct platform, planning various activities and content which can be posted, defining frequency of interaction, and strategizing methods to avoid page abuse. In this section, we report participants views and perceptions, which can help answer these questions. Section 5.1 and 5.2  discuss preferred OSM platform and audience for policing followed by discussion on various policing activities in Section 5.3. Next, we present analysis of major hindrances, apprehensions and vulnerabilities. 
We refer interviewed IPS officers as P1 \dots, P20, and citizens as C1 \dots, C17. 
\subsection{Exposure, Need and Audience} 
\textbf{\emph{Police exposure to OSM:}} We found that use of OSM for policing activities was majorly unexplored. Five officers reported that they were planning to adopt OSM and realized that it was a powerful medium. P1 said: \emph{``It is a very powerful medium you will have to be very conscious about how to use it.''}. However, 4 officers said that they were not planning to use OSM, as it was not needed at present. 
Among 8 officers who mentioned using OSM for official purposes, only 2 officers had used OSM for more than 2 years. We found Citizens were informed about police plans and actions on OSM. Ten citizens were aware of police pages; among these, 6 citizens had heard or seen police pages on OSM but were not following them and 3 citizens mentioned that they had visited these pages to communicate with police. Seven citizens were not aware of police pages.

\textbf{\emph{Information easily found on OSM:}} We asked officers if some information could only be found on OSM, and was difficult to find on other networks. Eleven officers felt that information like friend list, real time content, public mood, and opinion of young population were available on OSM. P3 said: \emph{``Youth is sometimes available only on OSM, what they have to say is sometimes not available on traditional media. They have very strong ideas which are otherwise missed out''} and P9 said: \emph{``young people are using OSM on mobile devices. They can upload real time content.''} Some officers (N=4) felt that it was not that information was available only on OSM but it could now be easily accessible on OSM. P15 said: \emph{``Some information might be easily available in OSM than searching it anywhere else. For e.g. I can get initial clues using OSM, if I have a target suspect and I want to find his friends.''} Eight citizens believed OSM use could ease information flow between citizens and police. C7 said: \emph{``Suppose there is threat in Delhi and they [Police] give me alerts on it [OSM] then I would have liked to follow it.''} Five citizens said that it could be difficult for police to use OSM. C5 said: \emph{``What I feel is police departments are busy with murder cases, they have to roam around places, they don't get time to check this status.''} When we asked citizens, if they would like to follow these pages, 11 participants gave positive response. However, 5 citizens refused to follow these pages, as they found them inactive or feared that police will monitor everything. C12 said: \emph{``Tough, I wont follow usually. If police is there, they will try to monitor everything. Lets say if I hold a bottle of beer on road they might come and arrest me. My privacy is lost.''}

\textbf{\emph{Target Audience:}} Strategy for effectively using OSM depends on the target audience of police pages~\cite{International-Association-of-Chiefs-of-Police:2013fk}. We found that there was no clear understanding of target audience on OSM among the officers. We asked officers, who according to them was the target audience they wanted to reach using OSM. They used ambiguous words like \emph{anyone available, general public, educated people, affluent classes, crowds assembling in different area or anybody resident of their area.} P6 said: \emph{``every citizen is our target audience, but audience is limited to account holders [on OSM].''} Six officers said that OSM was mainly for younger population and students. An officer said that it would depend upon the policing job assigned to him. 
\subsection{Use Facebook and WhatsApp for policing } 
Police need to select a platform that would suit their strategy and audience for effective OSM use~\cite{International-Association-of-Chiefs-of-Police:2013fk}. However, overabundance of such online / mobile based social networks makes platform selection a challenging task. We asked officers which OSM was most useful / helpful for law and order purposes and why they would prefer it. Officers preferred Facebook for policing activities and felt that it was more interactive than others. Six officers used Facebook, 7 officers used Facebook and Twitter, an officer used only YouTube and 3 officers mentioned using all 3 OSM. P4 said: \emph{``Facebook is more interesting, personal viewpoint can be expressed and its more interactive. People prefer to give their inputs through Facebook.''} Some officers believed that each platform had its own advantage. While comparing Twitter and Facebook, P7 said that Facebook was useful as it was a common platform and offered extensive reach, whereas Twitter was useful in spreading information or attracting feedback. Other factors, which influenced the choice of platform, were time constraints, size of network and kind of content.  P15 said: \emph {``Actually it is matter of how much time you can devote. FB gets too big and then you are involved with lot more people. Twitter you make a small comment and it is done.''}

\begin{figure}[!htb]\scriptsize
\captionsetup{font=scriptsize, labelfont=bf, textfont=bf}
\vspace{-1mm}
\centering
\includegraphics [scale=0.40] {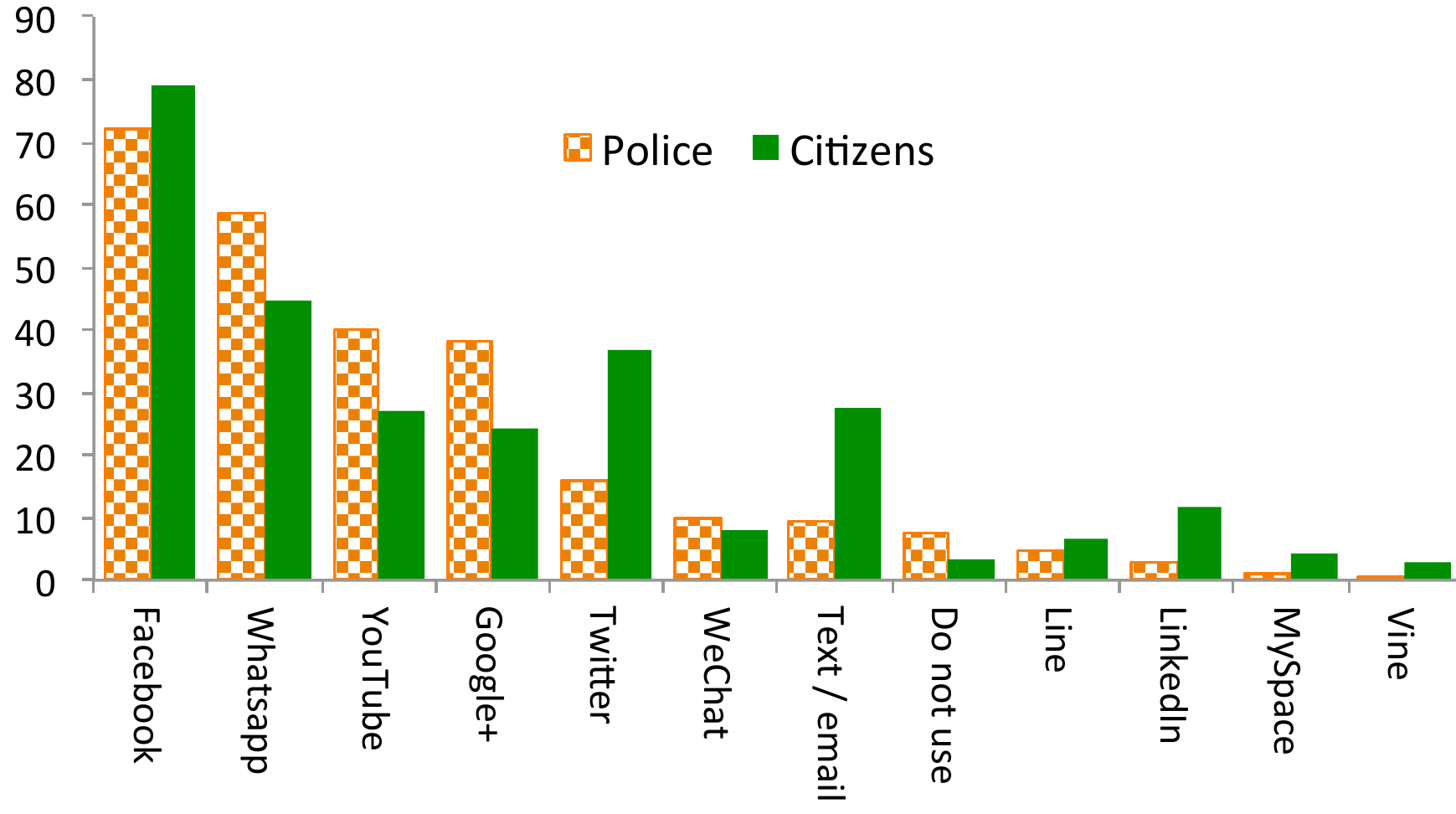}
\vspace{-3mm}
\caption{{Comparison of citizens (N=204) and police (N=445) choice for OSM as policing tool.}}
\label{fig:Facebookuse}
\vspace{-3mm}
\end{figure}

Officers also used mobile-based applications like WhatsApp, and BlackBerry Messenger but within their personal capacity. Ten officers felt that WhatsApp could be used to maintain law and order. Officers felt that instant image based communication on mobile phone, and notification that message was read was beneficial for police. P13 said: \emph{``During a gathering or law and order situation, a usual question is how many people [are there]. Then they [officers in control rooms] start guessing, if Officers on duty have a phone they can click a picture and upload it, so that in office we have an idea.''}  P12 mentioned that \emph{``Nobody can deny that they have received a message [Last seen feature].''} Few officers suggested using alternate telecom services to send bulk SMSs. According to them, too many channels (like Facebook, Twitter and WhatsApp) might confuse citizens and lead to loss of good quality information. To understand the mass opinion on preferred OSM, we asked police officers which OSM they would use for day to day activities. Survey results show that 72.17\% officers used Facebook. We asked citizens in survey that which OSM they would use to communicate with police. Almost 80\% citizens preferred to use Facebook, followed by Whatsapp (44.54\%). In comparison to 42.02\% citizens only 15.80\% policemen preferred using Twitter ($\chi^2$ test, p-value $<$ 0.01). Figure~\ref{fig:Facebookuse} shows comparison of preferred OSM by citizens and police. We found statistically significant difference police officers and citizens choice to use OSM ($\chi^2$ test, p-value $<$ 0.001). However most preferred platform was same for both. On Comparing existing studies, we found OSM preference of Indian police was similar to UK and the US police; all preferred Facebook.

\subsection{Policing activities on OSM -- Perceptions and Expectations}

We now discuss what police organizations hope to get from using online social media and citizens perceptions about how they would like police to use this media for better policing. OSM offers two-way communication facility, which can be used to build communities~\cite{Gupta-A.-Joshi-A.-and-Kumaraguru-P:2012cr}. However, we found that most officers preferred using it as a one-way-communication channel to either push information or pull information rather than as an interaction medium. Officers primarily considered OSM as an effective tool to monitor traffic problems, issue advisories, and to understand public opinions on various issues. The utility of these pages for investigation and monitoring was not very clear. However, citizens found OSM to be an effective tool for incidence reporting like eve teasing and neighbor-hood problems such as drunken people on streets. 

\subsubsection{Traffic Control -- Report offenders and jams}
Fifteen officers agreed that OSM could be an effective tool to issue advisories related to traffic conditions on roads, change in traffic routes, and traffic management. For instance, a cricket match of Indian Team and a popular Festival (Chhath Pooja) fell on the same day. Therefore to avoid traffic jams an advisory was issued through OSM, which proved helpful. Few officers also said that citizens could report traffic offences on OSM pages. He believed that before introduction of traffic policing through OSM, only traffic policemen on roads were responsible to catch traffic rule offenders but now anybody could upload a picture of traffic violations. Hence, OSM helped increase human resource, which could help solve traffic issues. However, few officers were not convinced that OSM could be useful for traffic policing. An officer remarked: \emph{``We started a traffic police page in Ahmedabad city. We started getting some feedback but we didn't know if it was popular. Results were limited, and were not up to our expectations. Therefore, it is not something which is going to be effective''. }

For citizens, traffic police pages were also useful to report issues like jams, accidents, beggars on roadsides, corruption like bribes to policemen, and issues with public transport. Eight citizens said that they would like to report traffic issues to police through OSM pages. A Female participant said: \emph{``I use public transport such as auto rickshaws; they [auto drivers] harass passengers a lot by either refusing to go or charging extra money. I will like to report that.''} C12 said that he could instantly geo-tag that location in the Facebook post, so that officers monitoring the page knew where exactly the incident happened and could take appropriate actions. C13 said that he had posted pictures of many traffic offenders on Delhi Traffic Police page and received an acknowledgement on the post that a ticket was issued to offender. However, he was not sure if the person was really issued a ticket, therefore lost interest in the complete exercise. Thus, participant was not convinced and felt need for a transparent system, which could guarantee punishment for offenders. We found no statistical difference between the two groups for use of OSM for traffic policing (p>0.05, Fishers exact test).

\subsubsection{Advisories}
Fourteen officers agreed to use OSM police pages for issuing advisories to citizens. Officers mentioned that it was a routine exercise to issue advisories on police websites and other media; similar practices could be adopted on OSM. We found that officers were interested in posting wide variety of advisories through OSM such as traffic arrangements, places to avoid during major events and festivals (``Police Bandobast''), crime alerts, women, children and senior citizens safety, VIP arrangements, and natural calamities like cyclone or floods. P2 appreciated the use of OSM for crime updates such as, \emph{``informing citizens about Chain snatching cases, time it could happen, precautions to be taken, they are on our website and can be given on OSM as well.''} 

\begin{figure}[!htb]\scriptsize
\captionsetup{font=scriptsize, labelfont=bf, textfont=bf}
\centering
\frame{{\includegraphics [scale=0.5] {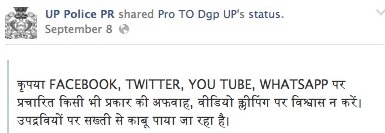}}}
\vspace{-3mm}
\caption{{Advisory shown to participants posted by UP police asking citizens not to trust any unknown video posted on YouTube, Facebook, Twitter and WhatsApp.}}
\label{fig:advisory }
\end{figure}

To develop better understanding about police and citizen perspectives about advisories issued through OSM pages, we showed them an advisory released on Facebook by UP state police. The advisory asked not to trust any rumors during Muzzafarnagar riots on OSM (See figure~\ref{fig:advisory }). Ten police officers and 8 citizens felt it was useful information and would like to repost such information to help it spread. An officer refused to share such a post as he regarded it as censoring communication. Five citizens said that this information as not useful or ambiguous to believe. C4 said: \emph{``I won't do anything. I don't think people who are in the riot area will actually have time to go through these.''} Some participants felt that this information could be fake and best would be to make ones own judgment. We asked the same scenario to citizens in online survey and found that 68.16\% people said that they would retweet or share this post, 18.99\% said that they would like it whereas 11.73\% survey participants said they would ignore the post.

\textbf{\emph{Cultural influence on advisory selection: }}We found influence of culture needs of people on OSM policing. Officers in different regions had different advisory requirement and priorities. For instance, P1 said that OSM could be used to spread awareness regarding social evil practices of Witch-hunting in northeast India.~\footnote{\scriptsize{Superstitious practice of declaring women as witches.}} An officer from West Bengal region said they would like to issue advisory informing them about regions where festival arrangements could be made (Durga Pooja -- largest festival in the region), and what forms they were required to fill to get approvals. An officer from Hissar, (rural area), in India responded that there was no communal problem in their region. Therefore, it was not useful for them to share advisory like communal riots in other parts of country. Officer from Haryana mentioned that they would like to create awareness women, children and senior citizen safety as one of the agenda.

\subsubsection{Investigation for corroborating evidences}
We found that 9 officers believed OSM was helpful investigations to obtain clues to solve cases as every activity on OSM left a trace. They found information like friend list, location, and photos was helpful in solving cases. P2 said that he had solved a case using this information.~\footnote{\scriptsize{Exact case not shared due to confidentiality reasons.}} Two officers said it could help get evidences in clueless murder cases, which can be used to build upon the case. P3 said: \emph{ ``Yes there have been cases of murder where this network and some patterns were found using OSM, obviously it helps. Very useful in criminal investigation.''} Two officers mentioned use of OSM in developed countries like the US (Boston bombing) and UK (London riots). P20 said that during London riots a citizen who posted his picture showing him vandalizing shops and wearing a stolen jacket was used in court of law against him. Eight officers felt that was a grey area and OSM role in investigations was yet not known but information about suspect / victim's network could be useful. Some officers also expressed concern that information shared on OSM during investigation might influence the case, cause witness harassment, and might even leads to media trial rather than constitutional justice. Two officers believed that effectives of OSM in investigations would depend upon the location of the servers. Getting permissions to get information from the servers located outside India takes time, which might lead to delay in investigation. Only 3 officers said that there was not much use of OSM in investigation process. P11 said: \emph{``It [OSM role] is close to zero for this [investigation]. Even if that information is their evidential value in court is almost zero.''} However, only 2 citizens mentioned that OSM could be used for investigations. While interviewing citizens, investigation did not come up as a possible use of OSM. However, 4 citizens expressed concerns about use of OSM for investigating cases. C12 mentioned that \emph{`` I don't think so, I don't think information in OSM is trust worthy therefore I am not sure if it can be used as an evidence not even secondary.''} We found statistical difference between number of  police and citizen participants in surveys, who marked crime investigations as a policing activity through OSM ($\chi^2$ test, p-value $<$0.05)

\textbf{\emph{Investigation a futuristic approach:}} We found that OSM was helpful in investigating cases but influence of OSM on investigation pace was not too significant at present. Seven officers believed that OSM did not have much influence on the pace of investigation and 4 officers believed that OSM role in investigation was dependent upon the type of case and was limited at present but could be useful in future. We found significant difference in police (M=1.89, SD=0.63, N=410) and citizen (M=2.26, SD=0.83, N=204) perceptions regarding influence of OSM on investigation pace (Wilcoxon rank-sum test, z=-8.78, p$<$0.001). In the survey, 87.31\% officers and 62.75\% citizens agreed or strongly agreed that Information obtained via online social media could help them solve investigations more quickly. To understand OSM policing in future, we asked survey participants to mark on a likert scale of 5, if they agreed / disagree that OSM in crime fighting / investigating activities will be critically important in the future. Almost 79\% citizens (M=2.01, SD=0.83, N=204) agreed / ~strongly~agreed with the statement whereas 94.51\% policemen (M=1.65, SD=0.65, N=419) agreed / strongly agreed with the statement (Wilcoxon rank-sum test, z= -10.2, p$<$0.001).

\begin{figure}[!htb]\scriptsize
\captionsetup{font=scriptsize, labelfont=bf, textfont=bf}
 \vspace{-4mm}
\centering
{\includegraphics [scale=0.19] {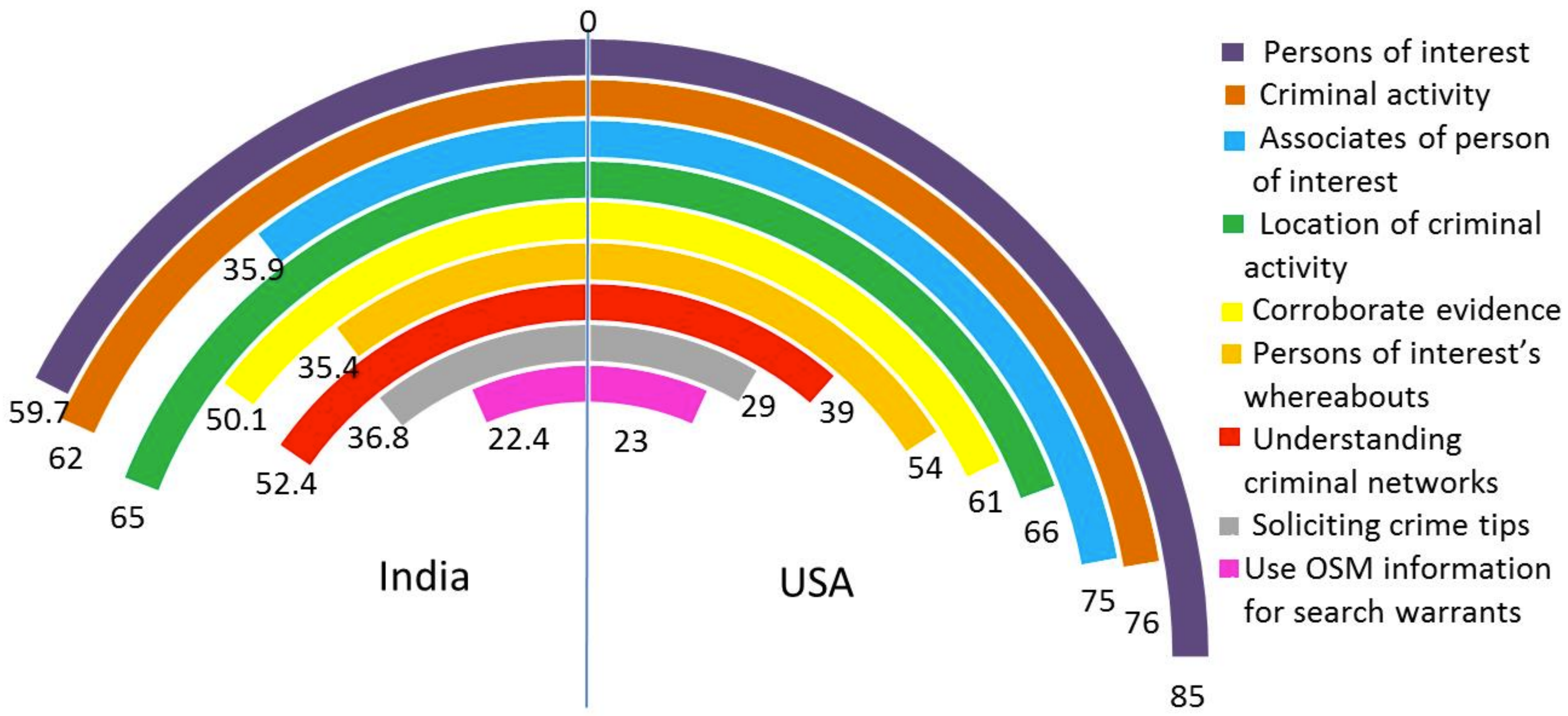}}
\vspace{-5mm}
\caption{ {
Comparison of Indian (N=445) and the US (N=1221) policemen choices for types of identification activity using OSM which help in investigating cases.}}
\label{fig:invest}
\vspace{-2mm}
\end{figure}

\textbf{\emph{Different countries -- different expectations:}} In surveys, we asked officers to mark different types of investigation that they could do on OSM.  More than 60\% officers preferred to identify criminal activity and location of criminal activity using OSM. Almost 60\% officers believed that OSM could be used to identify person of interests. More than 50\% believed that, OSM could help \emph{understand criminal networks and gather photos or statements to corroborate evidence.} We compared our results with expectation of US police in federal, state and local law enforcement agencies in the US published in a market research report and found statistically significant difference ($\chi^2$ test, p-value $<$ 0.001) between expectations of the two countries~\cite{Lexis-Nexis-Risk-Solutions.:2012uq}. We used the same question with some changes to suit Indian Context. Figure~\ref{fig:invest} shows officers' preferences for various investigative activities using OSM.

\subsubsection{Monitoring rumors and public opinion}
Eleven officers believed that OSM could be helpful in monitoring two prominent aspects  -- rumors that manifest in violent public upsurge and public opinion about various issues. Eight officers somewhat agreed that OSM could be potentially used for monitoring but mentioned various constraints like volume of content generated, and need for extensive cyber networks who could inform about events being organized on OSM. However, early information about rumors could be helpful in anticipating crime. 
Fifteen officers believed that rumormongers used OSM to incite crime and introduce panic among the citizens.  Among these many officers said that they would like to reduce panic and communicate the correct story using OSM. P1 mentioned about violence against northeastern Indians spread because of OSM. He said that people in Bangalore were made to believe [using OSM] that Assamese were discriminating on basis caste. Therefore, Bangalore citizens should also do the same with them. This made Assamese students to flee from rest of India back to Assam. Officers perceived that monitoring OSM could have helped understand what was agitating public mind and identify groups who radicalized people. 

Another need to monitor OSM was to understand public sentiments and common trends about various topics. Thirteen officers said that they would like to use online social media to understand public moods. This knowledge could help them strategize policing activities.  P1said: \emph{``Suppose in police we want to form an opinion about a topic in that case we can get a lot of information from different people. We can form an opinion / make policy decisions [based on this].} P3 said, \emph{`` When ever we think of OSM, we think it in terms of what are aspirations of public, what they want from us, what are their grievances and how we can address. OSM is more representative; police delivery can be more up to mark.''} Officers considered OSM to be a powerful tool for masses to express their views. We asked citizens if police could use OSM to understand public opinion on various issues. Only six citizens replied ``it was fine''. C6 said: \emph{``Your status speaks everything. They [Police] can know what is going in mind of citizens and take action accordingly.''} In surveys, we found statistical significant difference between citizens and police for monitoring OSM ($\chi^2$ test, p-value <0.05)

\vspace{-3mm}
\subsubsection{OSM for Feedback}
We analyzed police and public opinion on positive / negative feedback and abusive comments posted on police pages. We asked officers and citizens, what would be their reaction to a post in which citizens gave positive feedback to police e.g. I thank police for reducing crime in our city. We found that 10 officers said they would like to share the message. This would help boost the moral of police officers and would encourage them to get more feedback. An officer said that replying to each post might be difficult due to resource constraints. Six officers who refused to share the post believed that this would bind them to reply to negative posts. P10 said \emph{``if you respond to positive thing you should respond to negative posts too. Entering into any kind of debate on OSM may not be a good idea''}. In comparison to this, 9 citizens said they would like to spread / reply to such messages. In surveys, 64.29\% citizen said that they will like or favorite such a post and 30.77\% said that they would repost / share. We believe that police departments should consider replying and acknowledging positive feedback from citizens to help community grow and build trust. 

Next, we showed participants a post in which a citizen had complained that there was no police to help a girl being beaten on a highway. Eleven officers said that they would appreciate negative feedback and would like to take actions against such complaints.  P1 said: \emph{``We will check if actually it is happening and take action; subsequently will post about the action.''} Officers believed that OSM could help explain citizens the reasons for such lapse so that people can understand the situation better. Nine citizens said that they would share or like such a post so that police is informed about the issue and might let them understand the gravity of situation. A participant said that if many people complain about the same thing police might take some action. Two participants said that they might go to help the person in danger as police cannot be present everywhere. In surveys, we found that 52.20\% citizens said that they will share / retweet the post and 39.01\% said they will comment or reply to the post. This shows potential of OSM to understand disagreement expressed by the citizens to strategize better. 

Six officers expressed concerns regarding citizens' expression to communicate their thoughts to police. P10 said: \emph{``Expectation of people how police should use OSM is not based on clear understanding of law enforcement agencies. They think that the way they use informal style language on Facebook accounts law enforcement would also do the same. That is not possible.''} P14 and P17 experienced a lot of trash / abusive messages on existing helplines such as 100 [Emergency police number in India]. They felt that situation might be same with OSM pages, as mischievous individuals would be more interested in writing wrong things. An officer said that Twitter and Facebook wall had a potential to create havoc. Citizens also agreed that abusive words should not be used on these pages. Only 13.19\% citizens said that they would like or favorite the posts which are abusive to the police.

\subsubsection{Issues, citizen's would like to report} 
\begin{figure}[!htb]\scriptsize
\captionsetup{font=scriptsize, labelfont=bf, textfont=bf}
\centering
{\includegraphics [scale=0.4] {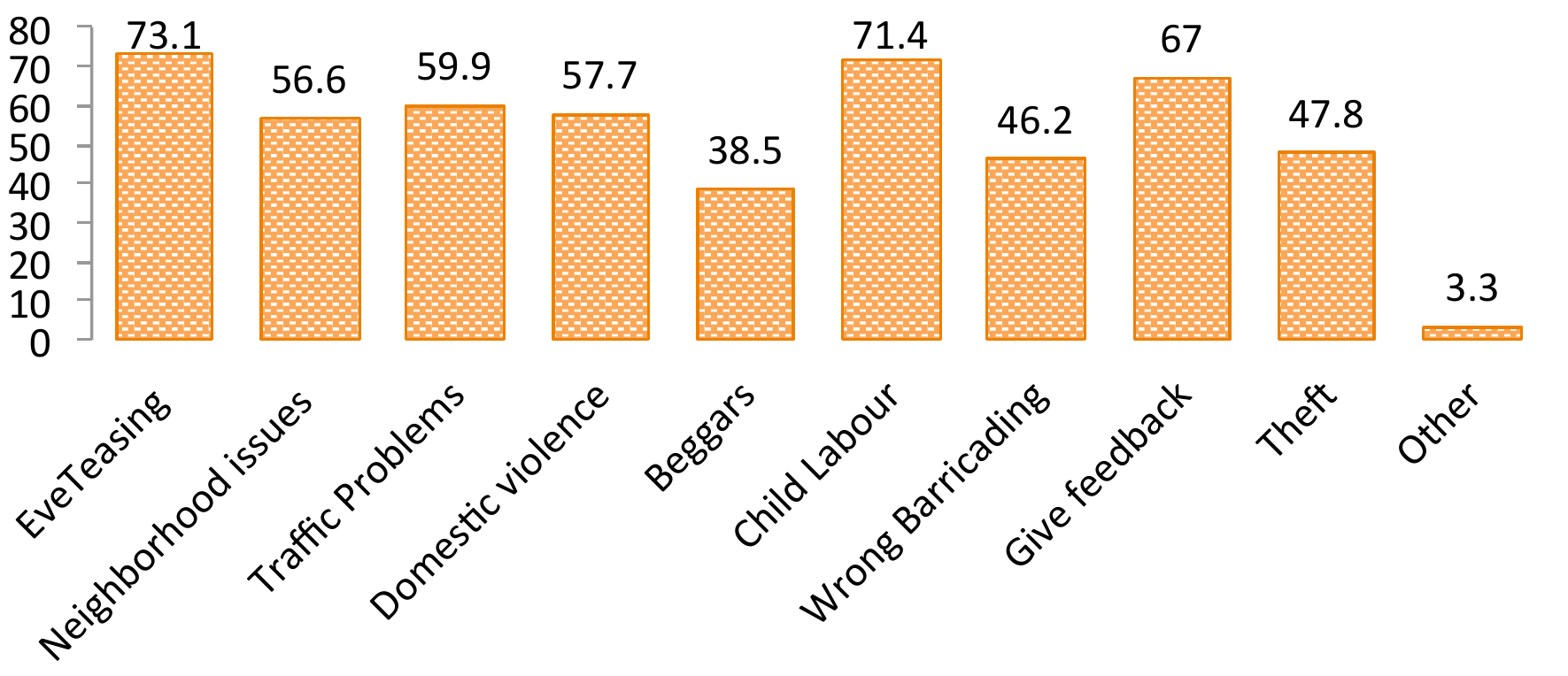}}
\vspace{-3mm}
\caption{{Citizen (N=204) issues which they will like to report using OSM.}}
\label{fig:citizenexpectations}
\vspace{-3mm}
\end{figure}

While discussing utility of OSM for policing, citizens mentioned various issues that they would like to report to police using OSM. 
Eight citizens said they would like to report eve teasing cases, 6 citizens mentioned about neighborhood issues and 2 citizens mentioned report cases of theft using OSM. C13 said: \emph{``if I see a girl in a vulnerable situation I might give this information, and they can take instant action''}. C16 said: \emph{``A friend of mine got 100 messages and multiple calls everyday from an unknown number. She filed a complaint but no action was taken. So she had to write to 10 other officers for help and finally she wrote to commissioner of police who helped her; But Facebook can make it easy.''} Citizens preferred reporting neighborhood issues, C7 said that he would like to complain against local Member of Legislative Assembly (MLA, politician), regarding bad conditions of roads, local shopkeeper in my area doing black marketing or if there are malicious people roaming around in street. He believed that through OSM anonymous complaints could be made against influential people, which was otherwise not possible. Figure~\ref{fig:citizenexpectations} shows issues which citizens will like to report using OSM.

\subsubsection{OSM improves coordination and prevents crime}
Fourteen officers believed that OSM could help improve citizen -- police coordination and prevent crime. Officers said OSM could bridge the gap between police and public. An officer said that if people knew that a senior officer reads these pages, they will be interested to use these pages. In real world it might not happen, as people might be hesitant to come forward. OSM could be a game changer if used in the correct spirit and would lead to a lot of information sharing. P14 said: \emph{``Yes, communication is meant for this [improved coordination], main problem is somehow communication needs to be there. Because broken people feel reluctant to come to us, additional communication channel is always helpful.''} Officers also felt that OSM was a helpful in crime prevention and would supplement real world actions such as police awareness programs. P1 said: \emph{``Yes, ragging problem is colleges, we do seminars but if we could give this information using OSM might help prevent such cases.''} We found no statistical difference between citizens and police preference for using OSM in crime prevention ($\chi^2$ test, p-value $>$ 0.05)

\subsubsection{Policing activities -- comparison with the US} 
\begin{figure}[!htb]\scriptsize
\captionsetup{font=scriptsize, labelfont=bf, textfont=bf}
\vspace{-3mm}
\centering
{\includegraphics [scale=0.33] {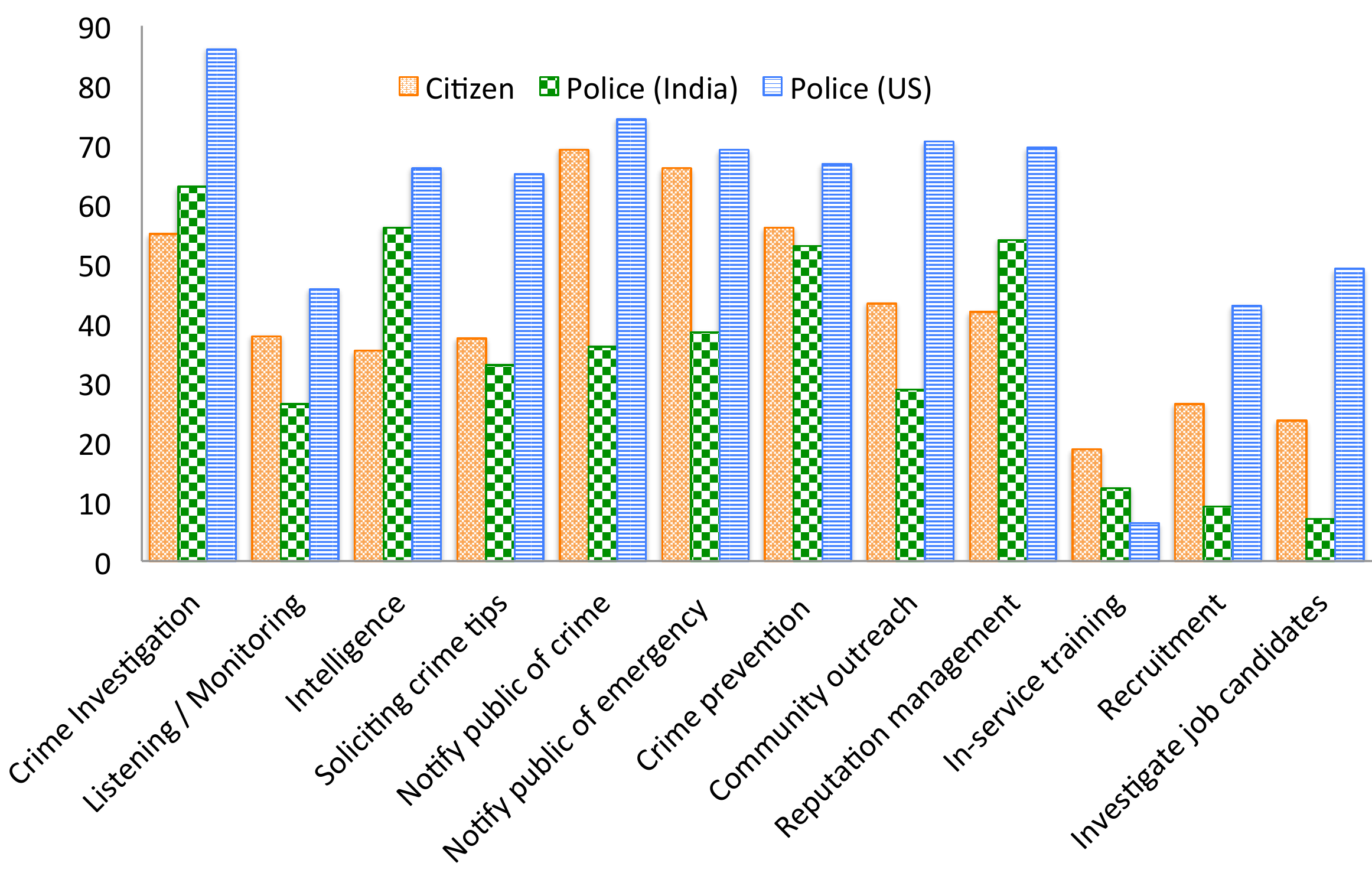}}
\vspace{-4mm}
\caption{{Choices of Indian citizens (N=204), policemen (N=445) and the US law enforcement executives (N=500) for policing activities on OSM.}}
\label{fig:compare}\vspace{-3mm}
\end{figure}

In surveys, we asked participants (both citizens and police) to choose activities that police should perform through OSM. We found significant difference between citizens and police expectations to use OSM for policing activities ($\chi^2$ test, p-value $<$ 0.001), however both preferred to use OSM for one-way communication. Only 28.67\% police departments considered using OSM for community outreach or citizen engagement, in comparison to 43\% citizens. Top three activities for which citizens thought police should use OSM were -- notifying the public of crime problems, notifying the public of emergency situation or disaster related issues, and crime prevention activities. In contrast to this, top three activities which policemen preferred were crime investigation, intelligence, and public relation / reputation management. On comparing police perception in India and the US, we found that crime investigation was most preferred activity for both police departments. However, in contrast to 36.1\% policemen in India, 63.7\% officers in the US choose to notify public of crime problems using OSM. Community outreach / citizen engagement was third most popular use of OSM among police officers (61.8\%) in the US in comparison to 28.67\% policemen in India. Figure~\ref{fig:compare} shows comparison of activities which police can perform using OSM as selected by citizens and police in India, and police in US. We found statistical difference between the US and Indian police for preferred policing activities  ($\chi^2$ test, p-value $<$ 0.001).

\subsection{Perceived constraints, and challenges}
In this section, we explore perceptions of officers and citizens regarding various constraints, limitations, and apprehensions that limit the successful use OSM for policing.

\subsubsection{Reach is 10\% only!}
According to 18 officers, reach of OSM was a crucial factor for successful use of OSM for policing. Most officers appreciated use of OSM for various policing activities; however, they believed that current penetration of OSM was much less than needed for successful use of OSM. Therefore were not sure if their efforts on OSM for improved policing would get good response. P6 felt that \emph{``As of now it will not be of great potential. Population of India having access to OSM would be 3 or 4\%.''}  
P7 said that OSM will be more helpful once it is available in local / regional languages. Few officers believed that large citizen participation on OSM could help get representative and unadulterated views of masses about various topics where as another officer believed that trustworthiness of information might reduce as accessibility goes high. P1 said: \emph{``large number of people of OSM we can get a good representative sample''}. During discussions with citizens, reach did not come up as a constraint. Citizens believed that OSM use for policing could be useful because social media was accessible to everyone. Few citizen participants believed that the Internet penetration in India was very low and therefor OSM might not be helpful for police. C5 said that only 5 -10\% people were connected to Facebook rest even did not know about Facebook and if any body wants to communicate with police on OSM they should have an account. C5 and C13 believed that not many people had the Internet connectivity on phone.

\textbf{\emph{Suitable for urban population:}} Some officers also believed that currently OSM was restricted to some urban areas and its utility was not much in rural areas. P9 said: \emph{``Urban sector, yes we can find some information. Rural areas, it is not of much priority.''} We found similar views among the officers regarding use of mobile phones application like WhatsApp; P8 believed that use of WhatsApp was not needed for next 20 to 30 years for policing but might be needed in urban area. P13 believed that WhatsApp users were not many and very few people had mobile phones in India. According to him rural people could not even SMS properly. Consequently, efforts to use OSM for policing were limited to some agencies in urban areas.

\subsubsection{Trustworthy information -- finding a pin in a haystack }
We asked participant that \emph{how trustworthy was the information obtained from online social media}. Thirteen officers believed that information obtained from OSM could be trustworthy depending upon the source of information, number of responses and would need human intelligence for verification. However, rest seven officers did not think that information of OSM was trustworthy but agreed that still police would need to verify any information is on OSM. Citizens also said that information obtained from OSM was not completely trustworthy. Six citizen participants said that information was not at all trustworthy or was difficult to trust whereas, 9 citizen participants sometimes trusted the information on OSM. Rest of the citizen believed that if people are expressing their grief on OSM, it should be trusted. Despite this, most citizens in interviews said that if information comes from police pages, it would be trustworthy.

\textbf{\emph{More response better trustworthiness: }}Among the 13 IPS officers who believed that information obtained from OSM could be trustworthy, six IPS officers felt that trustworthiness would depend on \emph{how many responses police could check }. P3 said \emph{``More are responses that we check more we are close to truth.''} They believed that mass participation on OSM added to the credibility of content. However, 2 officers (from rest seven officers who thought that information of OSM was not trustworthy) believed that very few people participated on OSM; therefore nothing could be commented about trustworthiness of this information. Three citizens believed that around 70\% of the information obtained via OSM was trustworthy whereas 2 citizen participants found information OSM to be 50\% times trustworthy.

\textbf{\emph{Source defines trustworthiness: }}Four officers believed that trustworthiness of information would depend upon the source of the information. P14 said \emph{``Trustworthiness of OSM [information] will depend on source as there is no instant verification of the source; information is not trustworthy.''} An officer believed that information coming from an organization could be trusted. We found that 6 citizens relied on the source of information to judge if it was trustworthy. C12 said \emph{``I don't trust the information, I verify the source of it.''} C6 believed that source could be trusted only if it is a verified account and said \emph{``Even police account can be fake. You never know. If police account is also giving information, I don't know how much can I trust it, we are living in a world where there fake profiles is a common thing.''} 

\textbf{\emph{Verification needed for deciding trustworthiness: }}Six officers believed that information obtained from OSM was a starting point and required human verification. P17 said: \emph{``Any information is just information as long as it is there. After that it has to be verified, correlated with other information, for this piece of information to go to the category of intelligence.''} P8 felt that finding trustworthy information on OSM was like \emph{``finding a pin in a haystack''} and needed cross verification from officers before it could be trusted. Three citizens said that they will cross verify the information; C13 said: \emph{``It is very difficult to rely on the information. For e.g. even leading journalist don't read in detail. It is important to read / cross check the original statement.''} However, a citizen said that there was no proof whether the information obtained from OSM was trustworthy.

\textbf{\emph{Trustworthiness -- Comparing to US:}} We found in interviews that both citizens and IPS officers found to difficult to trust information from OSM. However, in surveys, we found that policemen trusted information more on OSM to be trust worthier than citizens. In surveys we asked both citizens and police to mark on a likert scale of five, how much was the information obtained via online social media is trustworthy. In comparison to 49.63\% policemen (M=2.4, SD=0.84, N=391), only 17.03\% (M=3.15, SD=0.9, N=204) citizen survey participants agreed with statement (Wilcoxon rank-sum test, z=-1.19, p$>$0.05). We found similar to Indian police(58.06\%), high percentage of police officers in US (40\%, N=500) believed that information obtained through OSM was credible. Twenty six percent US (N=500) officers did not find information from OSM to be credible whereas only 12.27\% policemen in India disagreed or strongly disagreed that information from OSM was not trustworthy.

\vspace{-1mm}
\subsubsection{Fake profiles and anonymity}
We found that police need fake / pseudonym profiles for intelligence building, and investigation. Seven officers said that people might change behavior in case if they knew that they were interacting with police officer. P8 said: \emph{``Recently we saw in Holland some people made fake identities to catch hold of people involved in pedophile activities. Fake profile is needed to track crime and criminals.''} P13 associated fake profiles with undercover police concept and thought that intelligence agencies might require them. Eight officers felt fake profiles were not necessary for policing because these might risk police reputation, and also involved legal issues. P11 said: \emph{``it is not clear what legal issues it will create. Our reputation is at stake.''} P14 said that was right that fake accounts could be used to monitor but this was not how police works. \emph{``This is going bazar. There is some information somewhere we start. Once you have identified a person then a fake account is used.''} Few IPS officers considered it a controversial topic and refused to comment anything.

Another officer said that police informers might not feel comfortable revealing their identity on Facebook pages. To understand this behavior, we asked citizens if they would share personal information while communicating with police on OSM pages. Ten citizens said that they would prefer anonymous platform, whereas only 3 participants said that they would disclose the profile. Rest of the citizens believed that minimal information such as name or email id, should be sufficient. C8 said: \emph{``I can provide my e-mail id but not phone no., not any personal information''} A citizen said that he was fearful and concerned about his own security therefore used fake-profiles. C16 said: \emph{``Everybody can see it [post], I don't want others to know about that it's my thing, like in a police station it's between me and police. I might inbox message to police''}. Two officers also expressed concerns about citizens posting sensitive information, which could compromise their security. P20 shared an incident where geo -- location shared on OSM lead to theft in some foreign country, as thieves were aware when and for how long the house was deserted. He said that he always asked his family members not to share such information on OSM. To understand mass opinion about anonymity we asked citizens in survey --\emph{``How would you like to communicate information (e.g. complaints and feedbacks) to police using Online Social Media?''} We found that 36.26\% citizens would like to post anonymous information on police page whereas 29.67\% said that they will reveal minimal personal information e.g. email id. However, only 2.75\% said that they might create a fake account.

\subsubsection{Technical capability and interaction frequency}
Thirteen officers believed that police departments required trained people and special staff in police departments to leverage benefits of OSM. P2 said:  \emph{``[Person] who knows the limitation, has good knowledge about OSM.''} P9 said: \emph{``Team of better people as of now it's a resource constraint for us who can help us use it better.''} Officers felt team managing OSM should be acquainted with police decisions and very specific about content they post on OSM. P3 said: \emph{``Person who can be very specific about what he speaks, for now he will give very limited type of information.''} However, it was not clear who should be the appropriate authority to communicate with citizens. Some officers believed that dedicated staff could handle police pages under supervision of higher officials. Six officers believed that senior police officers, special cyber crime cells or technical service departments who maintained webpages could interact with public on OSM. 
Eight officers mentioned different kind of tools to use OSM for policing. Two officers among these, mentioned need  sentiment analysis tools and tools to know public opinion. Two officers said that they would need tools to trace back malicious people who spread misinformation. An officer said: \emph{``it was difficult to get diagnostics for Facebook as departments had been using to log email details. We need to contact them [Facebook].''} Two officers said that they would need systems, which can combine all tweets related to a particular subject and generate reports. An officer said that such tools were used in Boston bombing to extract information. 

\begin{figure}[!htb]\scriptsize
\captionsetup{font=scriptsize, labelfont=bf, textfont=bf}
\vspace{-3mm}
\centering
{\includegraphics [scale=0.42] {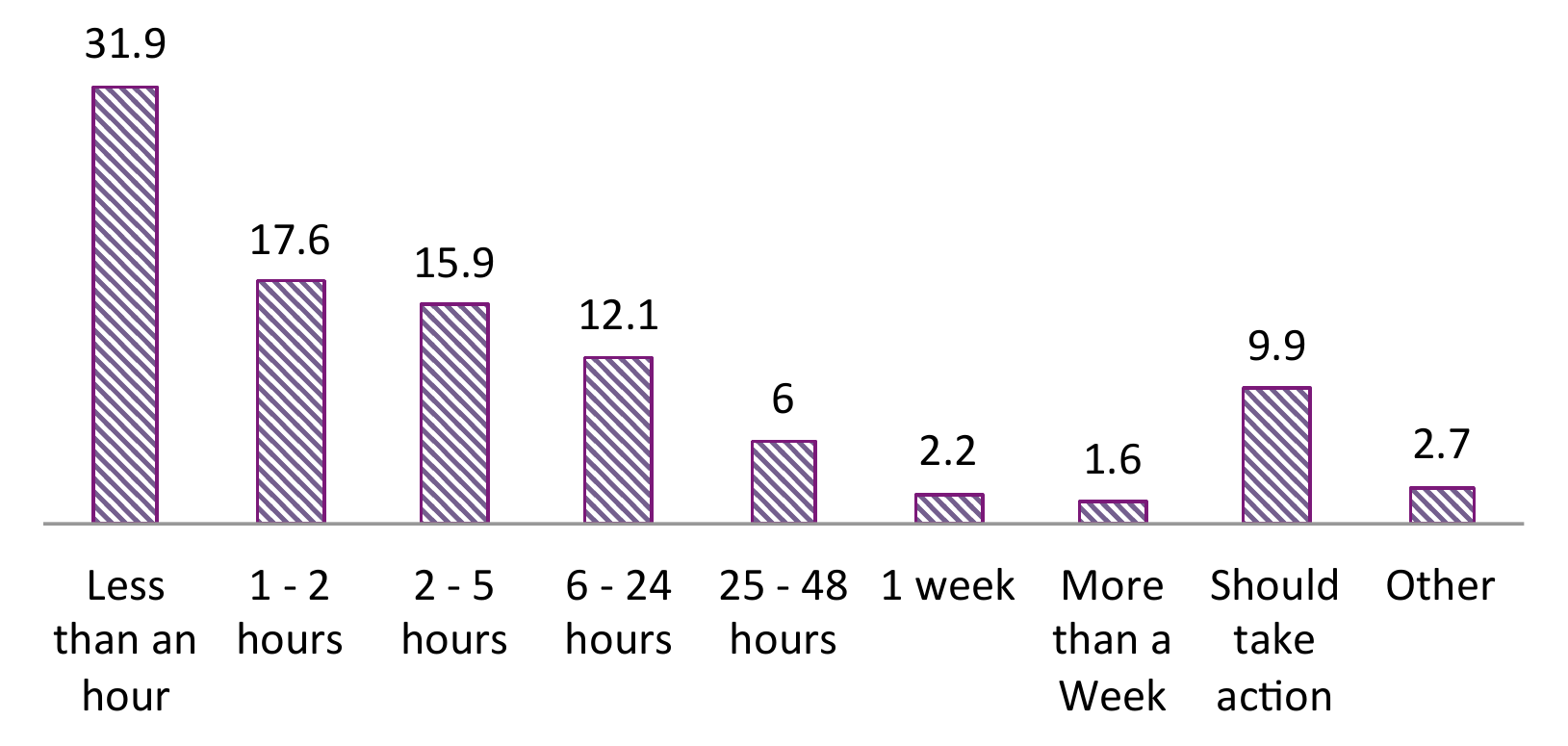}}
\vspace{-3mm}
\caption{{Citizens' responses (N=204) for time that police could take to acknowledge the citizens' post.}}
\label{fig:interaction}
\vspace{-3mm}
\end{figure}

Studies suggest that consumers expect quick response when they post complaints or request on OSM~\cite{kelly:2014fk}. We analyzed citizens' view on how long police could take to respond back to citizens' requests. Citizens' expectation varied from an hour to a week; three citizens said that police should take few minutes or less than an hour. C5 said: \emph{``if police has to use OSM, then there should be a team who should be checking it every second.''} Five citizens said that it would depend on the kind of complaints and volume of complaints received on OSM. C14 said: \emph{``That depends upon the issue, traffic problems could be answered in 1 or 2 days. Children / people who need immediate help like accident cases; should get immediate response.''} Two citizens said that police should send an immediate acknowledgement but it might take time to respond back with a solution. For rest of the citizens, respond time varied from 2 to 4 days. Contrary to citizens' expectations, none of the officers said that response could be provided in an hour or immediately. 
In surveys we found that 31.87\% said that police should take less than an hour to acknowledge that they had seen the post /message. Figure~\ref{fig:interaction} shows survey response of citizens for time police should take to acknowledge the post / message. 

\subsubsection{Need Guidelines \& Policies for OSM}
Majority of IPS officers said that there were no existing guidelines or specific policies for using OSM. Three officers said that there were some kinds of guidelines as police code of conduct, internal policies, or social media policy for government employees which they could refer to, if needed. All officers felt requirement for policies, which could provide structure to policing on OSM, and some guidelines for content to be shared. Officers believed that policies could help leverage benefits of OSM and also define boundaries for it. P5 said: \emph{``Centralized guidelines should be there on how to use Facebook, Twitter and YouTube else everybody will start using it in their own ways because it is boundary less thing; anybody can use anything from anywhere. To set some boundaries [policies] will be good. Structure needs to be created, what all to use, for what purpose, Twitter for what."} P6 said that policy would help decide the content, which could keep OSM pages live during lean periods otherwise community might loose interest. He remarked that \emph{``As a policy we need to have library of content about issues, projects that are available ready to be pushed so that each individual unit can use it whenever needed.''} Few officers also expressed concerns for servers being outside the country and need for policy to decide the extent for using OSM. P4 said: \emph{``It [OSM for policing] is still in experimental phase but to leverage its benefits policy has to be built in keeping in mind -- policy of usage, pros and cons. Like Facebook servers are outside India so how information (confidential) can be saved on them.''} P19 said that they might identify person of interest on OSM but would need help from servers, which was not easy. Few officers felt that there were no existing policies and would be very hard to define policies. P11 said: \emph{`` No formal guidelines and procedure for PRO are available even in current scenario. It is surprising that it is trial and error; history of organization helps define rules.''}  In surveys, we found that 34.46\% policemen believed that their organization did not have a written social media policy whereas 69.4\% of agencies surveyed in the US said they have a social media policy~\cite{IACP:2013kx}. We found statistical difference between Indian citizens (M=2.03, SD=0.84, N=204) and police (M=1.79, SD=0.58, N=399) on necessity of policies for effective use of OSM (Wilcoxon rank-sum test, z=-7.54, p$<$0.001). More than 75\% citizens and almost 94\% policemen felt need for policies (rules and regulations) for using and benefitting from OSM.

\subsubsection{Potential Barriers - comparing the US}
\begin{figure}[!htb]\scriptsize
\captionsetup{font=scriptsize, labelfont=bf, textfont=bf}
\vspace{-3mm}
\centering
{\includegraphics [scale=0.46] {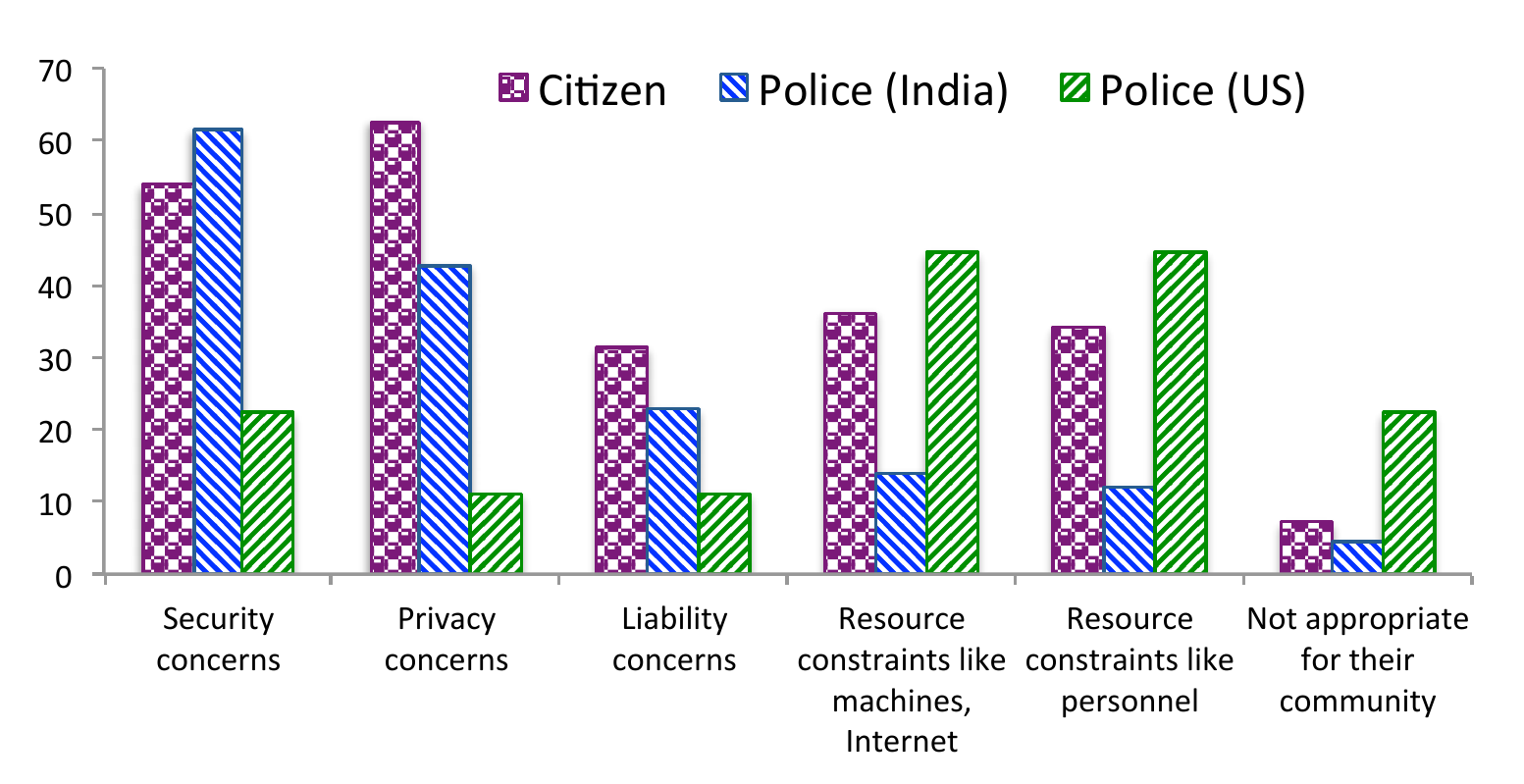}}
\vspace{-3mm}
\caption{{Citizens (N=204) and policemen (N=445) response to potential barriers for policing on OSM.}}
\label{fig:potentialbarriers}
\vspace{-1mm}
\end{figure}

We compared survey response of citizens and policemen in India to those in the US~\cite{IACP:2013kx} to the question -- \emph{``what are the potential barriers for using OSM for policing?''}  Top concern of policemen in India was security issue, whereas of the US police was resource constraint; however top concern of citizens in India was privacy. Almost 42.7\% policemen said privacy issues were potential barrier, whereas 62.3\%	citizens marked privacy as potential barrier. However only 11.1\% law enforcement executives in the US considered privacy issues as potential barrier. Figure~\ref{fig:potentialbarriers} shows comparison potential barriers as marked by policemen in India, in the US and Indian citizens.

\vspace{-6mm}
\section{Discussion}
In this paper, we explored OSM use by police departments in India and developed understanding on various aspects such as behaviors, perception, interaction, and expectations regarding policing through OSM. We examined statuses from various police pages -- Delhi, Bangalore, Uttar Pradesh and Chennai to understand various dimensions of police interaction with citizens on OSM. This was followed up with 20 interviews of IPS officers and 17 interviews of citizens to understand decision rationales and expectation gaps between two stakeholders (police and citizens). Finally we conducted 445 policemen surveys and 204 citizen surveys to present contrasts between expectations of Indian citizens, policemen and police officers in developed countries. 

We found that both citizens and police preferred Facebook and WhatsApp for policing activities. However, this study was conducted before Facebook bought WhatsApp, it will be interesting to study influence of this merger on participants behavior and expectations. Police officers expressed concerns that malicious entities might use sabotage OSM pages or citizens might ask controversial questions, which might create problems for police departments. Similar to developed countries, major concern was security implications of using OSM for policing. There were differences between officers and citizens' use of OSM for policing. Major policing activities included traffic management, rumor detection, public opinion analysis, and investigations to collect evidences. However, citizens expected that OSM could be used to report cases of eve teasing, theft, and neighborhood problems like drunken people on road, and beggars on streets. Officers showed concerns that reach of OSM was only 10\% and contrary to recent events like Muzzafarnagar riots where violence spread in rural using OSM, most participants believed that use of OSM was restricted to urban states. However, with increased penetration of mobile phone in rural areas, it will be interesting to see which kind of OSM would be preferred by the citizens and officers in future. Officers were worried that citizens might post sensitive information on OSM comprising their own security whereas citizens felt that privacy issue was a major barrier for using OSM to communicate with police. 

\textbf{Implication for research and practice:}
Research presented in this paper brings forth challenges, practices, and efforts required to manage police organization presence on OSM. Work shows implication on all three parts of the enterprise -- people (police and citizens), process (policing activities and communication), and technology (OSM pages). 
{{\begin{itemize}
\vspace{-1mm}
\item{People:} Police departments would need to identify people with necessary skills and leadership, which could help sustain and grow online communities for policing. Awareness needs to be created among the citizens regarding misuse / threats that might emerge, if OSM is not used appropriately. Police and citizens should decide on some house keeping rules, which can help improve OSM interactions. \\
\vspace{-6mm}
\item{Process: }Almost all officers and many citizens mentioned need for technically efficient and trained teams, which could assist police on ground to take appropriate actions. OSM framework exclusively for policing needs to be developed, which would require legal experts, police officers and technologists to contribute together. This could help improve evidential value of information posted on OSM pages.
\vspace{-2mm}
\item{Technology: }We found that police required technologies, which could help them use OSM effectively. Some of these technologies included public opinion mining, rumor detection, and better editorial controls, which could warn citizens about security and privacy issues. 
\end{itemize}}}

Police officers said that images, geo-location, videos and legitimate information on OSM is vital for investigations. However, extracting and validating such information from OSM is a challenge for security researchers. Thus, OSM adoption by police is not only a technological problem and interaction problem which needs broader assessment of policy, system designs and cultural influence on policing. We hope our study would help develop better strategies and technologies.  We understand that this study has some limitations; citizen survey participants were mostly in the age group of 18 -- 24 years. Although majority of OSM user base in India consists of participants in this age group, they are not representative of the entire India's population. Therefore, we believe that our results would provide general understanding of perceptions and challenges that police departments might face while using OSM. Future work could also look into attitudes and differences among older age groups and compare needs among technical and non-technical users. Future work could explore perception on international laws and regulation, which could be developed to curb negative influence of OSM use in policing. Finally, novel, usable mechanisms are needed to educate police officers and citizens to provide them with visibility and control over information exchanged on OSM policing pages. We hope that this helps develop transparent and trustworthy policing.

\vspace{-2mm}
\section{Acknowledgments}
We would like to thanks all participants for sharing their views with us and others who helped in data collection. Our special thanks to Mr. Nandkumar Sarvade who helped us connect with IPS officers.
We would like to thank all IPS officers who gave their precious inputs and shared their thoughts with us. We would like to thank all the members of CERC and PreCog who have given us continued support throughout the project; special thanks to Aditi Gupta, Siddhartha Asthana, Deepansha Sachdeva and Anuradha Gupta. 

%
\bibliographystyle{abbrv}
\small
\bibliography{osmpolicing}  
%
%
\appendix
\section{Appendix A}
Due to page limit, we could not include all the reference material here. The study protocols are available at: \url{http://bit.ly/1igvaiE} . The link contains both interview script and survey protocol. Citizen interviews.pdf is the script used for citizen interviews and IPS officer interviews.pdf is the script used for officers' interview. Citizen Survey.pdf and Policemen Survey.pdf contain the citizen survey and questionnaire and policemen survey questionnaire respectively.

\end{document}